\documentclass[twocolumn]{aastex63}

\newcommand{\msun}{\hbox{M$_{\odot}$}}

\newcommand{\re}{\hbox{${\rm R}_{\rm e}$}}
\newcommand{\mbh}{$M_\bullet$}

\received{\today}
\revised{}
\accepted{}
\submitjournal{ApJ Letters}

\shorttitle{Star formation in massive halos}
\shortauthors{Mart\'in-Navarro, Burchett \& Mezcua}

\begin{document}

\title{Quantifying the effect of black hole feedback from the central galaxy on the satellite populations of groups and clusters}

\correspondingauthor{I. Mart\'in-Navarro} 
\email{imartin@mpia.de}

\author[0000-0003-4266-5580]{I. Mart\'in-Navarro}
\affiliation{Max-Planck Institut f\"ur Astronomie, Konigstuhl 17, D-69117 Heidelberg, Germany}
\affiliation{University of California Santa Cruz, 1156 High Street, Santa Cruz, CA 95064, USA}

\author{Joseph N. Burchett}
\affiliation{University of California Santa Cruz, 1156 High Street, Santa Cruz, CA 95064, USA}

\author{Mar Mezcua}
\affiliation{Institute of Space Sciences (ICE, CSCIC), Campus UAB, Carrer de Can Magrans, 08193, Barcelona, Spain}
\affiliation{Institut d'Estudis Espacials de Catalunya (IEEC), C/ Gran Capit\`{a}, 08034 Barcelona, Spain}



\begin{abstract}

Super-massive black holes are fundamental ingredients in our theoretical understanding of galaxy formation. They are likely the only sources energetic enough to regulate star formation within massive dark matter halos, but observational evidence of this process remains elusive. The effect of black hole feedback is expected to be a strong function of halo mass, and galaxy groups and clusters are among the most massive structures in the Universe. At fixed halo mass, we find an enhanced fraction of quiescent satellite galaxies and a hotter X-ray intragroup and intracluster medium in those groups and clusters hosting more massive black holes in their centers. These results indicate that black hole feedback makes quenching processes more efficient through a cumulative heating of the gaseous intragroup and intracluster medium. 

\end{abstract}

\keywords{galaxies: clusters: general --- galaxies: clusters: intracluster medium --- galaxies: formation --- galaxies: evolution --- galaxies: star formation }


\section{Introduction} \label{sec:intro}

In the local Universe, every massive dark matter halo, and therefore every massive galaxy, hosts a super-massive black hole in the center \citep{Kormendy13}. Super-massive black holes grow in mass mainly through gas accretion \citep{Croton06}, and the total energy released during this process is on the order of a few percent of the final black hole mass (\mbh). For black holes more massive than $\sim$10$^9$ solar masses (\msun), the energy released is similar to the binding energy of the host halo itself and can therefore potentially alter the evolution of galaxies residing within. In fact, in state-of-the-art cosmological numerical simulations, the energetic feedback radiated by growing black holes is responsible for reproducing the observed properties of massive galaxies \citep{Schaye15,Weinberger18}. Moreover, the effect of black hole feedback is expected to become increasingly important with increasing halo mass.

Galaxy groups and clusters are massive, gravitationally bound systems and feedback from the super-massive black hole at the center of the dark matter halo is expected to play a dominant role in regulating their internal thermodynamics. In particular, numerical simulations of galaxy clusters show that the energy injected by the central super-massive black hole into the intragroup /intracluster medium (IGM/ICM) is able to heat the existing gas, preventing the formation of cooling flows \citep{Churazov02,Martizzi19}. Moreover, this feedback is thought to be strong enough to actually regulate the star formation of satellites living within these massive ($M \gtrsim 10^{12}$ \msun) dark matter halos \citep{Dashyan19}. 

Observationally, the clearest indication of the effect of black hole feedback on cluster scales is the existence of large X-ray cavities likely sustained by the energetic input from the central black hole \citep{McNamara07,Fabian12}. However, an empirical assessment of whether black hole feedback actually regulates star formation has proven to be a challenging task. Arguably, the main observational difficulty arises from the fact that black hole feedback is highly non-linear, with time variations much shorter than those associated with the quenching processes of galaxies \citep{Hickox14}. The observed connection between the properties of central and satellite galaxies, known as {\it galaxy conformity} \citep{Weinmann06}, has also been proposed to be a consequence of the aforementioned black hole-regulated star formation of the satellite population \citep[e.g.][]{Kauffmann13}, although it may also arise from the hierarchical nature of a $\Lambda$-Cold Dark Matter ($\Lambda$-CDM) Universe \citep{Bray16}. Direct observational evidence of black hole-suppressed star formation therefore stands as a fundamental open question, motivating the development of this study. 

In this letter we explore how the properties of satellite galaxies in groups and clusters depend on the mass of the black hole of the central galaxy. We find that dark matter halos hosting more massive black holes in their centers exhibit an enhanced fraction of quiescent satellites and are able to sustain a hotter IGM/ICM. The outline of this work is as follows. Data are presented in \S~\ref{sec:data}. We describe our main metric, the \mbh--$M_\mathrm{halo}$ relation, in \S~\ref{sec:MhaloMbh}, and our main results are presented in \S~\ref{sec:resu}. Finally, these results are discussed in \S~\ref{sec:end}.

\section{Data} \label{sec:data}
We have based our analysis on a sample of 4,308 galaxy groups and clusters \citep{Tempel14} selected from the Sloan Digital Sky Survey \citep[SDSS, ][]{SDSS10}. An extensive discussion about the sample properties is provided in \citet{Tempel14} (see also \S~\ref{sec:match}). We briefly note that group and cluster halo masses and virial radii are measured by assuming that the system is in virial equilibrium. The quoted halo masses and sizes are then estimated using the velocity dispersion and the radial extent of the detected group and cluster members, assuming a Navarro-Frenk-White profile \citep{Navarro97}. The location of the most luminous galaxy (hereafter central galaxy) is also provided for each object in the catalog . In total, the original catalog contains 82,458 (flux-limited) groups and clusters, at a median redshift of 0.0864, expanding a range in halo mass from $\sim 10^{12}$ to $\sim 10^{14}$ \msun.

We complemented our sample with total star formation rate \citep[SFR, ][]{Brinchmann04} and stellar mass \citep{Kauffmann03,Salim07} measurements also based on the SDSS for each group and cluster member. We did not include in the analysis individual galaxies flagged out by the MPA-JHU group. Further details on this data set can be found on the MPA-JHU project website\footnote{\it https://wwwmpa.mpa-garching.mpg.de/SDSS/DR7}.

\section{The \mbh--$M_\mathrm{halo}$ \ relation} \label{sec:MhaloMbh}

\begin{figure} 
  \begin{center}
  \includegraphics[width=8cm]{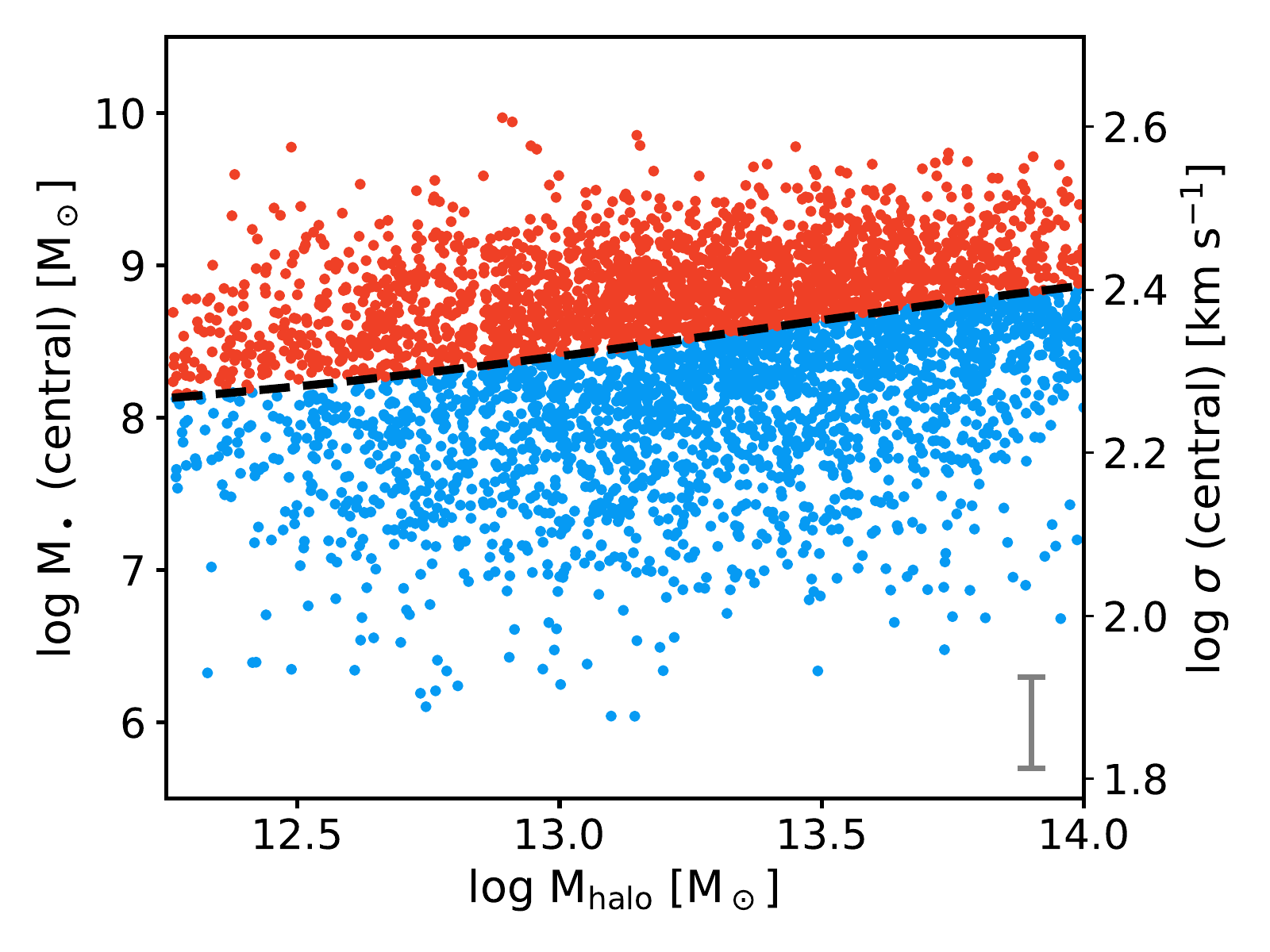}
  \caption{The vertical scatter across the median \mbh--$M_\mathrm{halo}$ \ relation (dashed lined) is a proxy for the total energy released by the central black hole into the IGM/ICM. Over-massive black hole halos (orange dots) have more massive black holes in their centers than under-massive black hole ones (blue dots) and, therefore, should have experienced more intense feedback processes. Each point in this figure corresponds to one of the 4,308 SDSS groups and clusters in our sample. Halo masses were calculated through a dynamical modelling of the satellites, while black hole masses were estimated by measuring the stellar velocity dispersion of the central galaxy. The typical uncertainty in the black hole mass estimation is shown on the bottom right corner.}
  \label{fig:1}
  \end{center}
\end{figure}

In order to explore the connection between central black hole feedback and star formation in satellite galaxies, we first constructed a relation between the group or cluster halo mass (M$_\mathrm{halo}$) and the mass of the black hole (\mbh) hosted by the central galaxy as shown in Fig.~\ref{fig:1}. Black hole masses were estimated using the \mbh-$\sigma_e$ relation presented in \citet{Remco16}, where stellar velocity dispersions correspond to a fixed aperture of 1~\re. 

For each central galaxy, the velocity dispersion and the redshift were measured from its SDSS optical spectrum using the pPXF algorithm \citep{ppxf} fed with the MILES stellar population synthesis models \citep{miles}. We then used the measured redshift to estimate the distance to the central galaxy, from which we derived the fraction of its effective radius covered by the SDSS fiber (R$_\mathrm{SDSS}$=1.5 arcsec). Finally, the stellar velocity dispersion at 1 R$_e$ was calculated by applying an aperture correction to the value directly measured from the spectra \citep{Cappellari06}.

Having halo and central black hole mass estimations, we calculated the average \mbh--$M_\mathrm{halo}$ \ relation using a two-step running median scheme. First, we measured the median central black hole mass over 150 equally log-spaced halo mass bins (from $\log$ M$_\mathrm{halo}/\msun =$ 12 \ to $\log$ M$_\mathrm{halo}/\msun=$ 14). Then, we fit these 150 median halo and central black hole masses with a low order polynomial. This allowed us to have a well-behaved M$_\bullet($M$_\mathrm{halo})$ function, which is shown in Fig.~1 as a black dashed line. The results presented in this paper are based on a 4th order polynomial for the M$_\bullet($M$_\mathrm{halo})$ function, but we also repeated the analysis with higher and lower polynomial orders, finding similar and consistent trends.

The \mbh--$M_\mathrm{halo}$ relation shown in Fig.~\ref{fig:1} allowed us to define a metric which is effectively sensitive to the amount of energy radiated by the central black hole \citep{MN16,Terrazas16}. We labeled as {\it over-massive} black hole halos those systems lying above the mean \mbh--$M_\mathrm{halo}$ relation. Complementarily, {\it under-massive} black hole halos are those that, at a given mass, have a central black hole less massive than the average. The reasoning behind this separation and its relation with the effect of black hole feedback is simple. Halo mass is thought to be the main parameter describing the properties of gravitationally bound structures in a $\Lambda$-CDM Universe, from individual galaxies to groups and clusters \citep{Blumenthal84}. Hence the (vertical) scatter in Fig.~\ref{fig:1} is expected to probe systems with very similar properties but with a variety of central black hole masses. Since the net energy released by super-massive black holes is expected to scale with their total mass from theoretical arguments, feedback-related processes such as galaxy quenching are likely enhanced in over-massive black hole halos compared to under-massive black hole ones. Thus the scatter in the \mbh--$M_\mathrm{halo}$ relation probes the effect of black hole feedback on group and cluster scales, modulo the assumption that over-massive and under-massive black hole halos only differ with respect to their central black hole masses.
 
\subsection{Homogenizing over-massive and under-massive black hole halos} \label{sec:match}
It is critical that our approach ensures that there are no systematic differences between over-massive and under-massive black hole halos other than the mass of the central black hole. This is needed to isolate the effect of black hole feedback from possibly confounding variables related to, e.g., cluster assembly \citep{Wechsler18,Bradshaw19}. In order to make sure that the over-massive and under-massive black hole halos are as indistinguishable as possible, we adopted the following strategy. For each group and cluster, we identified seven key properties that may potentially affect the measured SFR values: total halo mass, halo size, total stellar mass within the halo, average stellar mass of the lightest and heaviest satellites (10th and 90th percentile by mass, respectively), average stellar mass of those satellites close to the center of the halo (10th percentile in $R/R_\mathrm{vir}$), and average stellar mass of those satellite galaxies in the outskirts (90th percentile in $R/R_\mathrm{vir}$). Then, for each over-massive black hole halo we found the most similar under-massive black hole halo in that seven-dimensional parameter space using a Python implementation \citep{scipy} of a KD Tree algorithm \citep{Maneewongvatana99}. We finally rejected those systems where no successful match was found (i.e., with a normalized KD distance larger than 0.1), leading to a total of 4,308 massive halos as indicated above.

\subsection{Stellar velocity dispersion as a proxy for \mbh}

The use of stellar velocity dispersion as a proxy for black hole mass has been extensively used in galaxy samples where direct black hole mass measurements are not available \citep{Yu02,Marconi04,Benson07,Bluck14}. Furthermore, it has been shown that, at fixed stellar mass, varying black hole mass is approximately similar to varying velocity dispersion in relations with stellar population and SFR properties \citep{MN19}. These findings further support our use of the stellar velocity dispersion as a proxy for black hole mass in our sample. 

However, a few important caveats should be noted. First, at fixed stellar velocity dispersion, there are apparent differences in black hole mass \citep{MN16}, i.e., the intrinsic scatter in the \mbh-$\sigma$ relation is not zero \citep{Beifiori12}. Second, the relation between black hole mass and stellar velocity dispersion might change with galaxy mass, internal structure, and even orientation \citep{Xiao11,Graham13,MN18c,Sahu19}, which implies that trends with halo mass might be biased \citep{Bernardi07}. These uncertainties and systematics related to the use of the velocity dispersion as a proxy for black hole mass support our use of a running-median to separate over-massive and under-massive black hole halos, which in practice has no other meaning than differentiating the two populations, acknowledging our lack of information about the actual black hole masses. 

\section{Analysis and results} \label{sec:resu}

\begin{figure*}   
  \begin{center}
    \includegraphics[width=7.9cm]{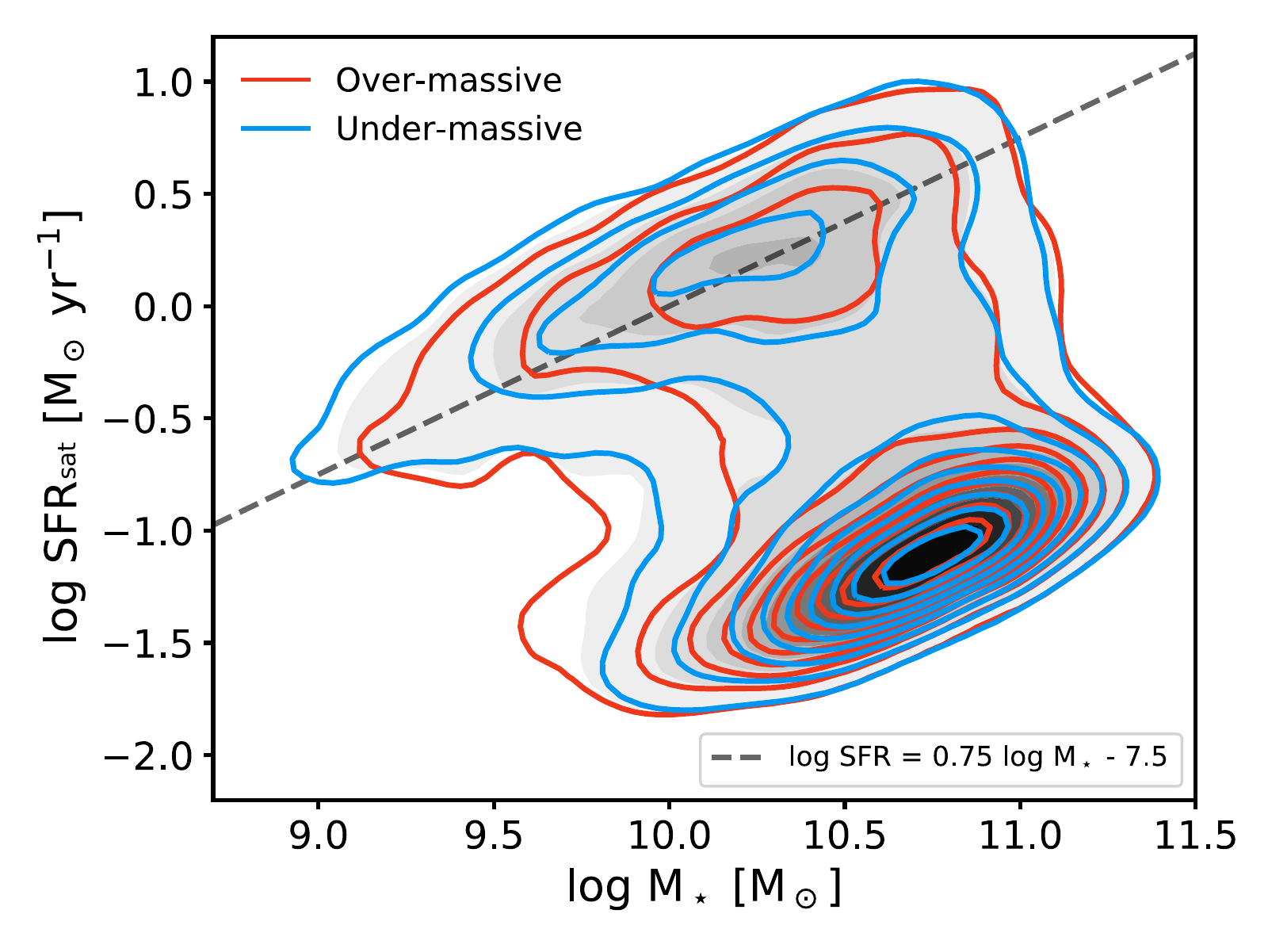}
    \includegraphics[width=7.9cm]{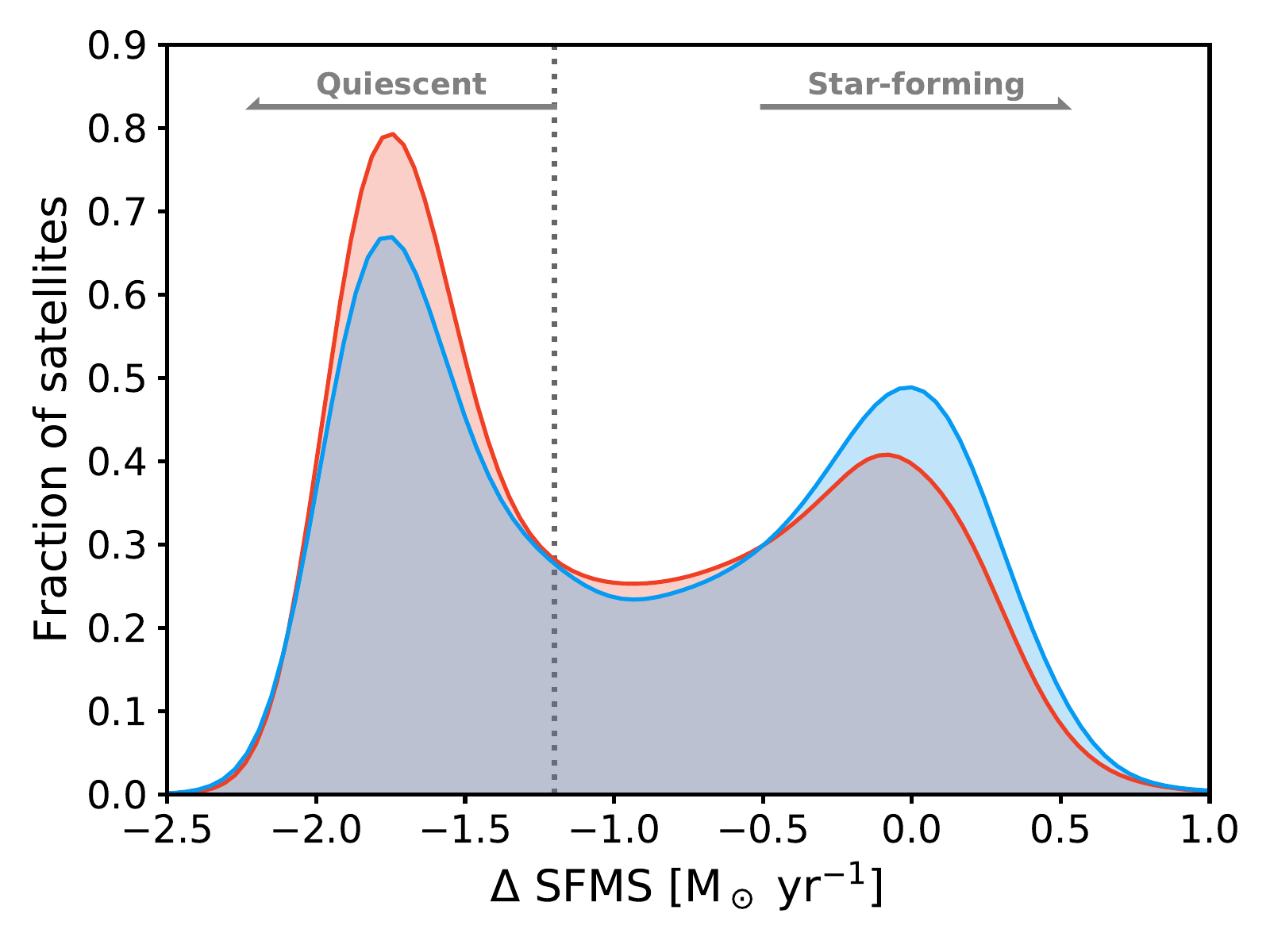}
  \caption{{\it Left.} SFR as a function of galaxy stellar mass for satellite galaxies in our sample. Regardless of the mass of the black hole, satellites follow a rather similar star formation main sequence (dashed line). {\it Right.} Distribution of distances (at fixed mass) with respect to the main sequence ($\Delta$ SFMS). The fraction of quiescent satellite galaxies ($\Delta SFMS < -1.2$, vertical dotted line) is different between over-massive (orange) and under-massive black hole halos (blue), as those groups and clusters with more massive black holes in their center tend to host an enhanced population of quenched satellite galaxies.}
  \label{fig:2}
  \end{center}
\end{figure*}

Using the $M_\mathrm{halo}$-\mbh \ relation as a local metric, we then analyzed the SFRs of satellite galaxies in over-massive and under-massive black hole halos. The left panel in Fig.~\ref{fig:2} shows that non-central members in both over-massive and under-massive black hole halos follow similar distributions in the SFR-stellar mass parameter space, with notably coinciding star formation main sequences (SFMS) highlighted by the dashed line\footnote{The SFMS shown in Fig.~\ref{fig:2} results from jointly fitting satellite galaxies in over-massive and under-massive black hole halos.}. However, a striking difference is revealed on the right panel when analyzing the distribution of distances from the main sequence ($\Delta$ SFMS): the population of quiescent satellites ($\Delta SFMS < -1.2$)\footnote{We set this threshold as a conservative limit defining the extension of the red sequence.} is enhanced in groups and clusters hosting more massive black holes in their centers. Because the number of satellite galaxies in over-massive and under-massive black hole halos is almost identical, by normalization, the number of star-forming galaxies is enhanced in the latter.

\begin{figure*}
  \begin{center}
    \includegraphics[width=7.9cm]{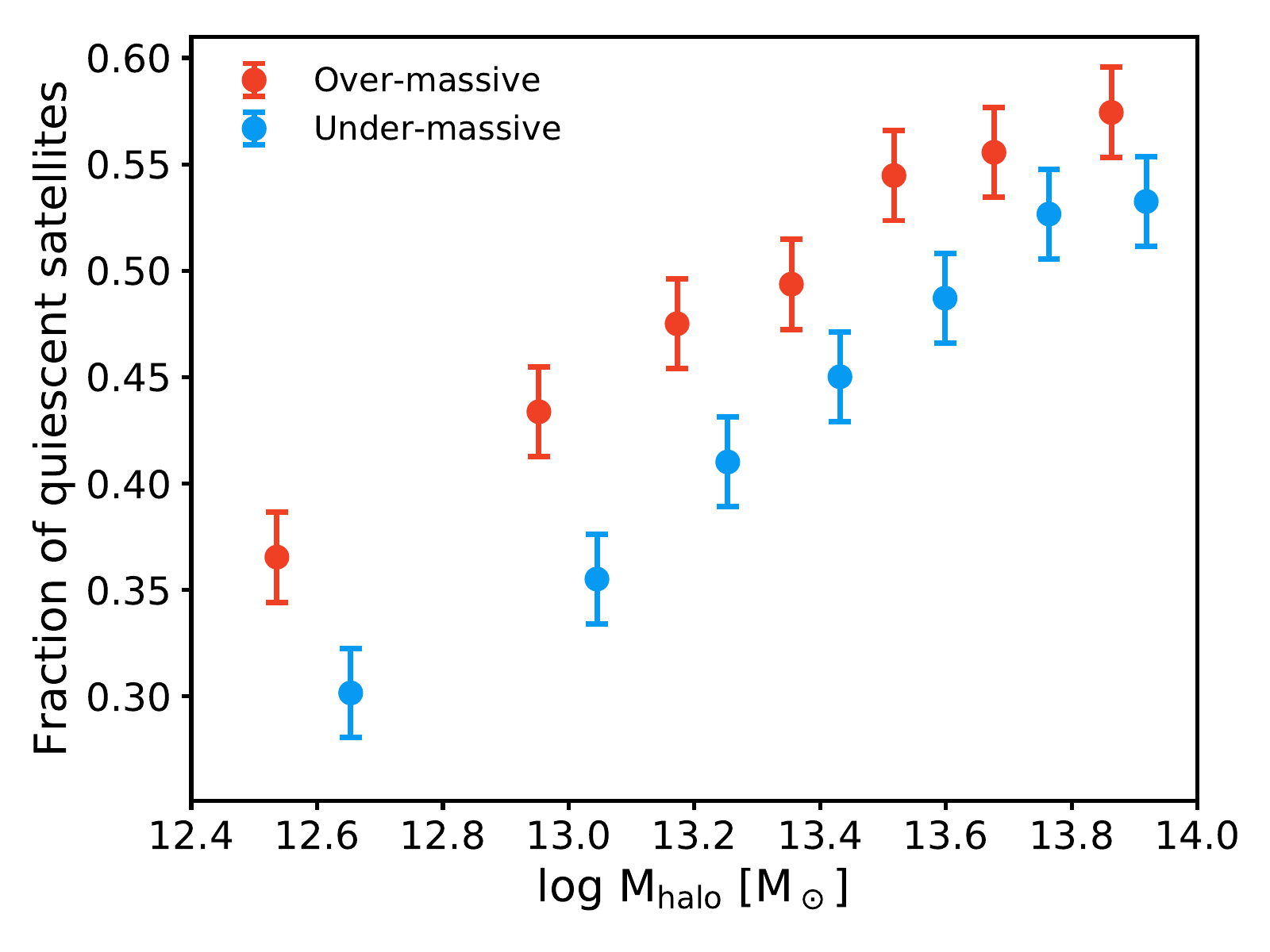}
    \includegraphics[width=7.9cm]{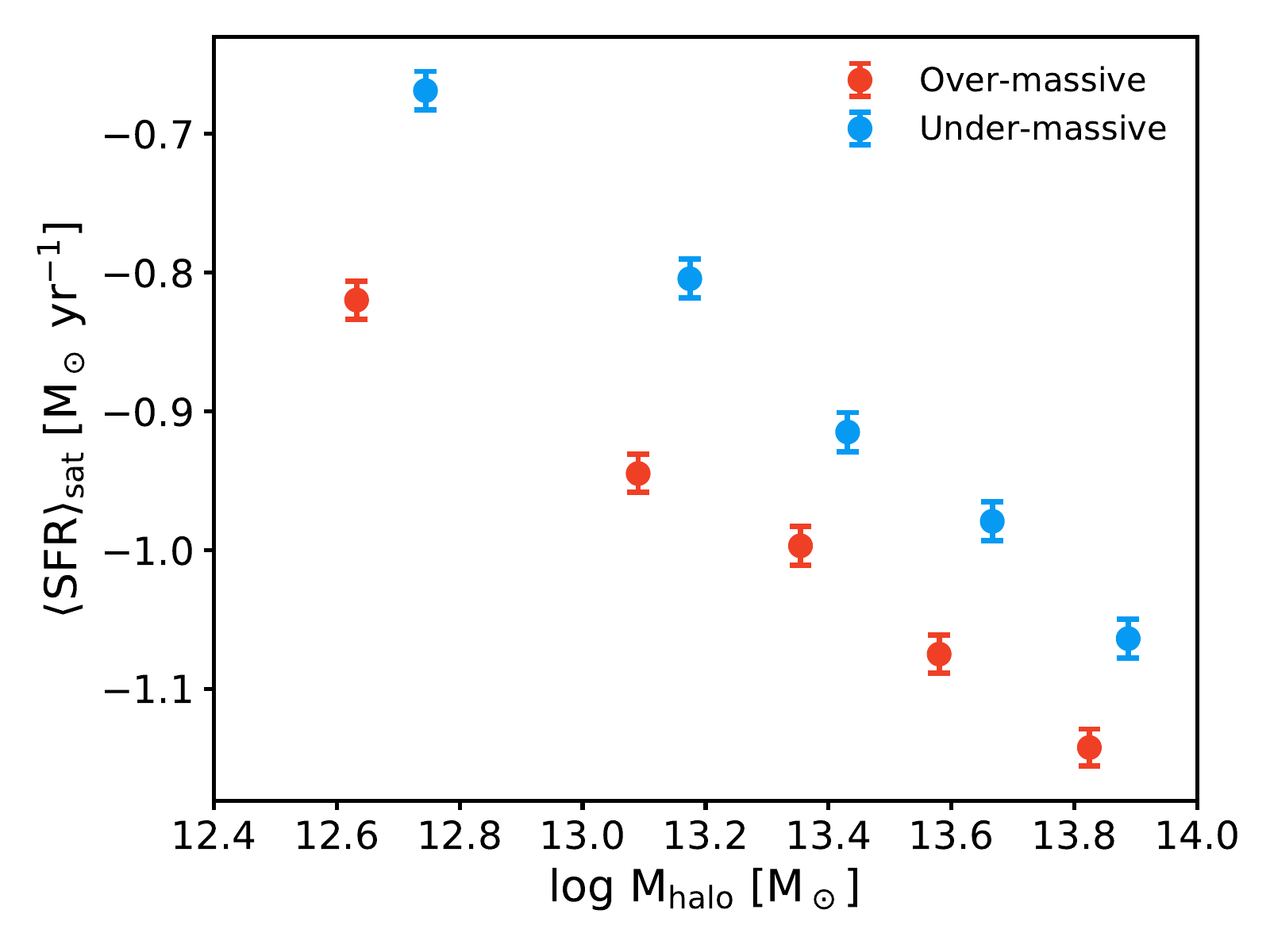}
  \caption{Orange and blue symbols correspond to over-massive and under-massive black hole halos, respectively. Error bars indicate the 1$\sigma$ uncertainty level. {\it Left.} The fraction of quiescent satellite galaxies ($\Delta SFMS < -1.2$) increases with halo mass, but it is always systematically enhanced in groups and clusters with over-massive black holes in their centers. {\it Right.} Conversely, the median SFR of groups and clusters decreases with increasing halo mass, but it is higher in under-massive black halos because they tend to host fewer quiescent members.}
  \label{fig:3}
  \end{center}
\end{figure*}

In order to investigate the dependence of this effect on halo mass, we binned galaxies in Fig.~\ref{fig:2} according to the masses of the halos to which they belong. We then calculated the fraction of quiescent satellite galaxies and their average SFR. Fig.~\ref{fig:3} shows these two quantities for over-massive and under-massive black hole halos as a function of halo mass. The fraction of quiescent galaxies increases and the average SFR decreases with increasing halo mass, as expected \citep{Wetzel13}. On top of these trends, the differences between over-massive and under-massive black hole halos are clear for all masses. At a given halo mass, when a cluster or a group hosts a more massive black hole in the center, the fraction of quiescent galaxies is enhanced and the average SFR is lower.

A final piece of evidence comes from the analysis of the X-ray properties of 196 optically-confirmed groups and clusters \citep{Takey13}. These systems were identified as extended X-ray sources with an optical galaxy over-density counterpart detected in the SDSS. Unfortunately, there was little overlap between this sample and that shown in Fig.~\ref{fig:1}, so we did not have a dynamical measurement of the halo mass. Thus, for the analysis of the 196 groups and clusters with X-ray measurements we used $M_{500}$, an estimate of the average halo mass within an aperture of R$_{500}$ (where the mean mass density is 500 times the critical density of the Universe at the halo redshift), rather than $M_\mathrm{halo}$.\footnote{For the 22 systems that did overlap between the two samples, we found an approximate conversion of log~$M_\mathrm{halo}$= log~$M_{500} - 0.35 \pm 0.11$.} This $M_{500}$ halo mass was calculated iteratively based on the X-ray bolometric flux assuming a $\beta$ model \citep[see details in][]{Takey11,Takey13}.

\begin{figure}
  \begin{center}
    \includegraphics[width=8cm]{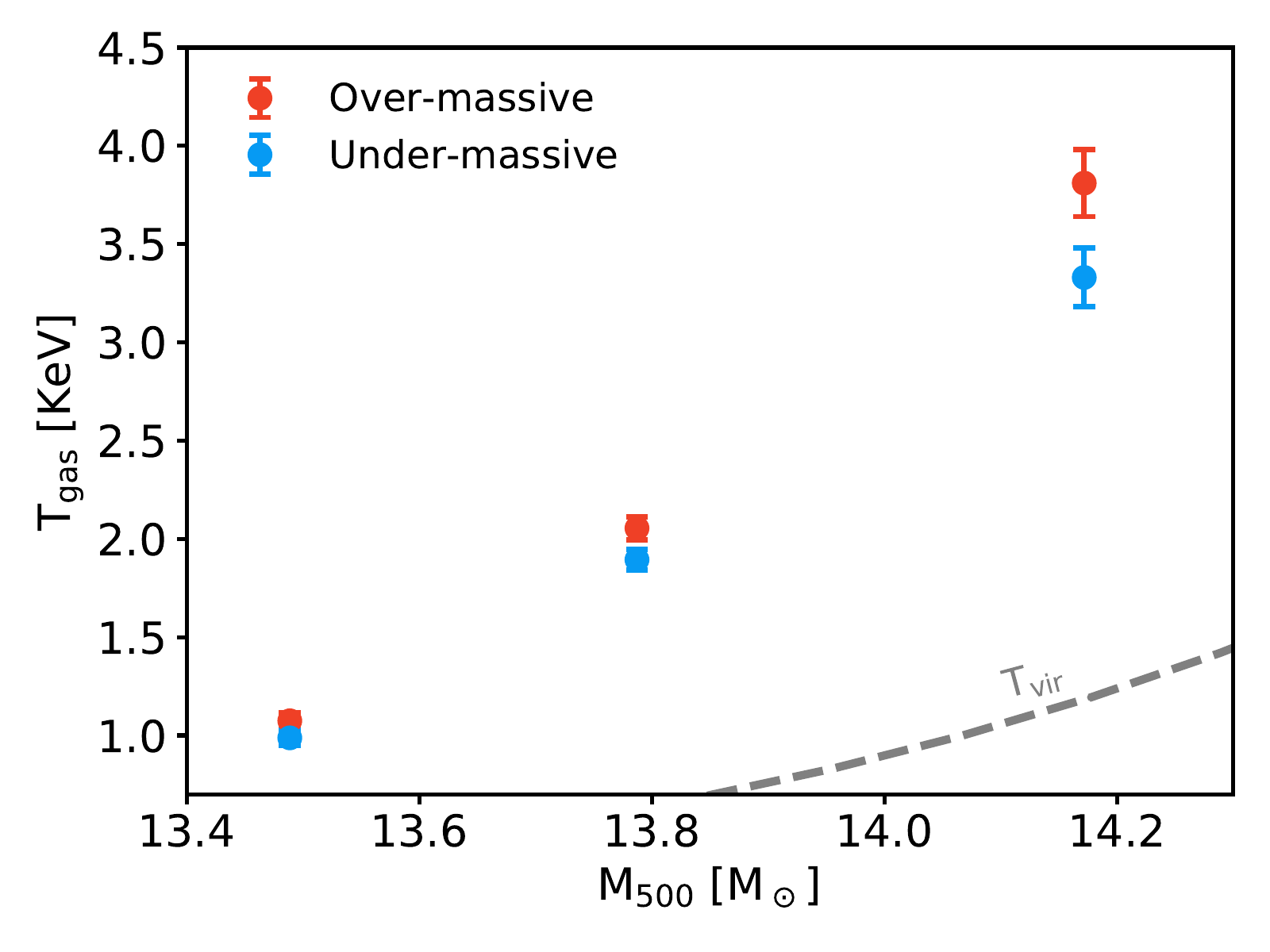}
  \caption{The X-ray gas temperature of the IGM/ICM increases in massive halos for both over-massive (orange symbols) and under-massive (blue symbols) black hole halos and in all mass bins the temperature is higher than the virial one (dashed line). However, on top of the halo dependence, there is a clear dependence on the mass of the central black hole. At fixed mass, halos with more massive black holes in the center sustain a systematically X-ray hotter IGM/ICM. Error bars indicate the 1$\sigma$ uncertainty level in the average temperature.}
  \end{center}
  \label{fig:4}
\end{figure}

For each of the 196 groups and clusters, the temperature of the IGM/ICM was derived from X-ray spectral fitting \citep{Takey11} and the mass of the central black hole was estimated in the same manner as before. As in Fig.~\ref{fig:1}, we divided the sample into over-massive and under-massive black hole halos and then calculated the average X-ray temperature in different mass bins. Fig.~\ref{fig:4} shows the $T_\mathrm{gas}$--$M_{500}$ relation for over-massive and under-massive black hole halos. Two features clearly emerge from this figure. First, as expected by virial arguments \citep{Birnboim03}, X-ray temperatures increase with increasing halo mass \citep{Bogdan18}. Second, the gas has been heated above the expected virial temperature, and this excess of energy correlates with the mass of the black hole in the center of the halo. At fixed mass, halos with more massive central black holes sustain hotter IGM/ICMs. 

\section{Discussion and conclusion} \label{sec:end}

Explaining the trends shown in Figs.~3 and 4 in the absence of feedback from the black hole in the center of the halos would require a highly fine-tuned mechanism. By construction, our over-massive and under-massive black hole halos are indistinguishable in mass, size, total stellar mass, and stellar mass distribution. The effect of a possible assembly bias \citep{Wechsler18} is likely negligible as there is little difference in the normalization of the SFMS between over-massive and under-massive black hole halos ($\Delta \ \mathrm{SFMS}_0 \sim 0.13 \pm 0.22$). Satellites in groups and clusters with more massive black holes form on average fewer stars, and there is an enhanced fraction of quiescent galaxies compared to under-massive black hole halos, suggesting that longer time-scale quenching processes such as strangulation may play a greater role in over-massive black hole halos. A hotter ICM would be more conducive to stripping the gaseous halos of infalling galaxies \citep{Joe18,Zinger18}, depriving them of their immediate reservoirs for star formation.

Moreover, given the departure from a simple scaling of the virial temperature (dashed line in Fig.~\ref{fig:4}), it is not trivial to explain the higher X-ray temperature of the IGM/ICM in those halos hosting more massive black holes in their centers without invoking feedback effects. Our analysis is certainly a simplified assessment of the effect of black hole feedback. We have assumed that the central galaxy is the only galaxy contributing to the heating of the IGM/ICM. Additionally, group and cluster galaxies likely formed under particular conditions in dense filaments even before falling into the group/cluster environment, and it is therefore possible that some differences between over-massive and under-massive black hole halos have been inherited. We note, however, that SFR measurements are sensitive to time-scales much shorter \citep{Calzetti13} than the dynamical time-scales within halos, and therefore they should be largely driven by recent IGM/ICM conditions.

Black hole heating acting on group and cluster scales emerges as a simple explanation to the observed trends, consistent with our cosmological understanding of galaxy formation \citep{Dashyan19}. If the temperatures of halos, and their internal thermodynamics in general, are set by a combination of virial plus black hole heating, quenching processes such as strangulation or ram pressure stripping would become more efficient in those groups and clusters with more massive central black holes, resulting in a higher number of quiescent galaxies as shown in Fig.~3. Moreover, this group/cluster-wide feedback process would emerge from accreting black holes in the center of individual massive galaxies, where similar trends between SFR, X-ray temperatures, and black hole masses have been reported \citep{MN18b,MN19}. Interestingly, a unified black hole heating scenario, from galactic to cluster scales, would naturally predict a certain degree of (one halo) galactic conformity \citep{Weinmann06,Kauffmann13}, as quenched galaxies are also found to host higher mass black holes \citep{Terrazas16}. Our results observationally support a scenario where the baryonic cycle in galaxies is regulated by the joint effects of black holes and dark matter halos across cosmic time. 

\acknowledgments

We would like to thank the anonymous referee for their useful and constructive comments on our manuscript. IMN acknowledges support from the AYA2016-77237-C3-1-P grant from the Spanish Ministry of Economy and Competitiveness (MINECO) and from the Marie Sk\l odowska-Curie Individual {\it SPanD} Fellowship 702607. M.M. acknowledges support from the Spanish Juan de la Cierva program (IJCI-2015-23944) and the Beatriu de Pinos fellowship (2017-BP-00114).

%






\bibliographystyle{aasjournal}



\end{document}